\documentclass[aps,prb,twocolumn]{revtex4}

\newcommand{\fsize}{0.45\textwidth}

\usepackage{amsmath}    
\usepackage{amssymb}
\usepackage{graphicx}   

\begin{document}

\title{A statistical mechanics model for free-for-all airplane passenger boarding}
\author{Jason H. Steffen}
 \affiliation{Fermilab Center for Particle Astrophysics, Batavia, IL 60510}
 \email{jsteffenATfnalDOTgov}
\date{\today}

\preprint{FERMILAB-PUB-08-068-A}

\begin{abstract}
I present and discuss a model for the free-for-all passenger boarding which is employed by some discount air carriers.  The model is based on the principles of statistical mechanics where each seat in the aircraft has an associated energy which reflects the preferences of the population of air travelers.  As each passenger enters the airplane they select their seats using Boltzmann statistics, proceed to that location, load their luggage, sit down, and the partition function seen by remaining passengers is modified to reflect this fact.  I discuss the various model parameters and make qualitative comparisons of this passenger boarding model with models which involve assigned seats.  This model can also be used to predict the probability that certain seats will be occupied at different times during the boarding process.  These results may be of value to industry professionals as a useful description of this boarding method.  However, it also has significant value as a pedagogical tool since it is a relatively unusual application of undergraduate level physics and it describes a situation with which many students and faculty may be familiar.
\end{abstract}

\maketitle

\section{Introduction}

The study of physics requires one to learn knowledge and acquire skills which can be applied to a host of different fields, many of which are not obviously related to physics.  Some realizations of this fact include the development of computer models to describe various scenarios, others may be in the methods employed to optimize ``real-world'' problems, such as the canonical traveling salesman problem that can be solved via simulated annealing\cite{metro,press} or entropy considerations in image reconstruction\cite{press}.  Others still might be in modeling the behavior of people or objects by the application of physical theory.  For example, \citet{bachmat} used the principles of General Relativity to model airplane passenger boarding.  They found that the boarding process can be described in the context of a two-dimensional Lorentzian geometry.  Here the time required for a passenger to sit in their assigned seat is related to the number of other passengers that lie in his or her future light cone.

In the vein of airplane boarding, a recent study\cite{steffen} applied a minimization algorithm that is commonly employed in physics (Markov Chain Monte Carlo) to optimize the passenger boarding process.  This is not the first attempt to optimize passenger boarding, indeed, computer simulations have often been employed to study the boarding process\cite{steffen,belgians,vandenbriel}.  Yet, all of these studies, like that of \citet{bachmat}, require that the passengers have assigned seats.  While most carriers assign each passenger a seat, not all do.  Some discount carriers employ a free-for-all (FFA) boarding approach where passengers can sit anywhere that they wish inside the aircraft.  This boarding strategy, while it may appear chaotic and prone to disaster, in truth gives among the fastest boarding times in the industry\cite{wired}.  This fact alone would merit a study to understand the underlying principles of the FFA boarding procedure.

A cursory attempt to model FFA boarding would be to assign each passenger a random seat within the aircraft.  However, there are subtleties that such a model ignores which are fundamental to the success of this boarding method.  Of particular importance is the fact that each passenger chooses the seat where he will sit based upon the state of the cabin at the time that he enters.  A passenger may prefer to sit in either a window seat or an aisle seat if those seats are available.  Or, they may prefer to sit near the front or rear of the airplane.  Finally, their preference may suddenly change based upon the decisions of the passengers who are in front of them in the line.  It is this capacity to adjust based upon the state of the airplane that makes the FFA boarding process so successful.

I claim that basic statistical mechanics provides an excellent framework with which to study FFA airplane boarding.  In this article I present a model for FFA boarding where each seat inside the cabin is assigned an ``energy'' which reflects the overall desirability of a particular seat.  Then, when a passenger enters the airplane he chooses which seat to occupy using Boltzmann statistics.  As the airplane fills, the partition function is updated to reflect the evolving state of the aircraft cabin.  Thus, the FFA boarding process can be studied as a sequence of decisions by the passengers; decisions modeled by statistical mechanics considerations.

In addition to studying the boarding process, one may study the outcome or final seating arrangement of the aircraft for flights which are only partially filled (the seating arrangements of full flights are trivial) or, similarly, the seating arrangement at any given time during the boarding process.  Here, instead of modeling the seat selection of each individual passenger during boarding, one uses statistical mechanics to determine the probability that a particular seat will be occupied by a passenger at the time of interest---treating the passengers as a fermion gas and using Fermi-Dirac statistics.

The discussion will proceed as follows.  In the next section I detail the model for the airplane and the statistical properties of the passengers.  I then discuss the application of the airplane and passenger model to the boarding process; outlining how passengers move within the airplane.  Following that, in section \ref{results}, I discuss the airplane boarding times that the model gives as well as the effects of changes to the model parameters (section \ref{parameters}).  Section \ref{compare} gives a brief comparison of the results of this model to the results of the assigned-seating model used in \citet{steffen}.  Section \ref{fermigas} changes focus and looks at the final states of a partially filled aircraft using Fermi-Dirac statistics to determine the occupancy of the seats at the end of the boarding process (or, presumably, at any instant during the boarding process).  Finally, I discuss the utility of this model for various classes of people including students, educators, and airline executives.

\section{Model Description}

The airplane model and algorithm that I use to board passengers has three distinct components: the geometry and characteristics of the airplane, the movements of the passengers inside the airplane, and the decision-making process of the passengers.  The first two components are largely technical details and are somewhat less important in terms of the ``physics'' of the model.  The third component, on the other hand, is where statistical mechanics plays its major role; the mind of the typical passenger is modeled therewith.  Below I discuss these different components, note however, that some of the details of the airplane geometry and passenger movement could be implemented differently without affecting the important conclusions of this work.

\subsection{Decisions, energy, and temperature}

In this model, each seat in the airplane is assigned an energy which characterizes how desirable that seat appears to the average passenger.  If a seat has a small energy or a large negative energy, then it is more likely to be chosen (I use negative energies here solely for the image of having the passengers ``bind'' to the seats---adding a constant to the seat energies will not affect the results).  When a passenger enters the airplane, they choose their seat with probability $\exp(-\epsilon_i/T)/Z$ where $\epsilon_i$ is the energy of seat $i$, $T$ is the temperature, where
\begin{equation}
Z = \sum_i e^{-\epsilon_i/T}
\label{partition}
\end{equation}
is the partition function, and where the counter $i$ is over all accessible seats.  Here Boltzmann's constant ($k_{\text{B}}$) is set to unity so that temperature is measured in units of energy; I use electron Volts (eV) for simplicity, though the choice is arbitrary.  To choose a seat, a random number between 0 and 1 is generated and the Boltzmann factors are summed until the total probability is greater than the random number.  The last seat to be added to the sum is the selected seat.  Note that since these decisions are made by individual passengers based upon the currently available seats, they are analogous to a single particle choosing from all available single-particle states; thus the sum of the Boltzmann factors (\ref{partition}) is the appropriate partition function.

The temperature of the passengers is a measure of their apathy (or, perhaps good-naturedness) in the seat selection process.  At very high temperatures ($T \gg | \epsilon |$ where $\epsilon$ represents the typical size of the energies of the seats) all of the seats are essentially equal, the energies being a very small fraction of the ``thermal'' energy of the passengers when they make their seat selection.  At very low temperatures, the differences between seats appear large and passengers choose the best available seat at all times.  Presumably the temperature is determined both by human nature and by the environment surrounding the passengers (perhaps including things ranging from the length of the security line and the traffic leading to the airport to the quality of the jokes told by flight attendants and whether the musak reminds the passengers of ``good times'').  While most passengers are not likely to be at precisely the same temperature because of their different histories, using a representative temperature to characterize the mean behavior should be appropriate, particularly if the boarding process is reproducible across several different flights.  Given the consistent performance of FFA boarding\cite{wired}, this assumption is likely to be justified.

\subsection{Airplane description}

Like the study described in \citet{steffen}, the nominal (fiducial) airplane model used for this study seats 120 passengers and has six passengers per row with 20 rows.  There is no first-class cabin.  The energies of the seats are symmetric about the aisle such that a window, middle, or aisle seat for a given row has the same energy as its counterpart across the aisle.  In addition, there is a linear trend in the energies from the front to the back of the airplane.  Initially, this trend serves to make the front of the plane more desirable than the back.  Later in this paper (section \ref{parameters}) I investigate changes to this, and other parameters.  Having the energies of the seats vary with higher-order polynomials instead of simply a line may be a more accurate description, but as with virtually all parameters for this model it would need to be calibrated empirically with passenger data---an exersize that is beyond the scope of this work.

\begin{figure}
\begin{center}
\includegraphics[width=\fsize]{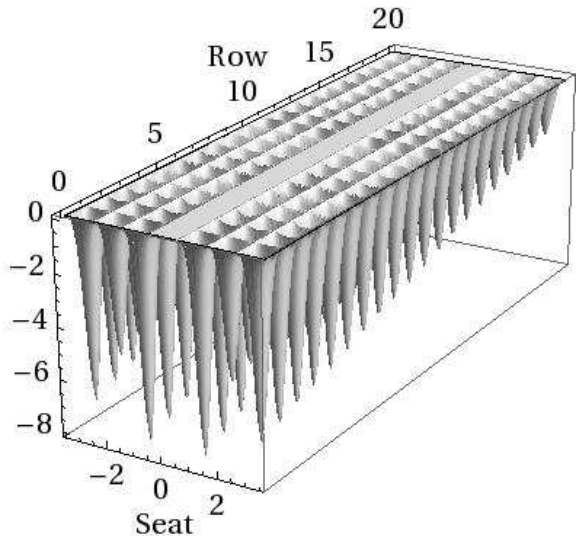}
\caption{A plot of the seat energies for the nominal airplane model.  The depth of the potential well for each seat corresponds to the probability that it will be selected by a passenger; the deeper the well, the more likely it will be chosen.  A constant offset can be added to these energies without affecting the results.}
\label{seatmap}
\end{center}
\end{figure}

The three energies for the seats in a given row are nominally assigned values of $-8$eV for aisle seats, $-7$eV for window seats, and $-5$eV for middle seats and the temperature is fixed at unity.  The effect of these energies is that the aisle seats are chosen 70\% of the time, the window seats are chosen 26\% of the time, and the middle seats are chosen roughly 4\% of the time.  The linear trend is also nominally selected such that the energies in each successive row increases by 0.25eV.  The result of this choice is that the probability of the first passenger sitting in the last row of the airplane is roughly 1\% of the probability that he will sit in the first row.  Figure \ref{seatmap} shows the energies of the seats inside the aircraft.

\subsection{Passenger movement}

For the boarding process, each passenger is assigned an amount of time that they require to load their luggage, a random number between 0 and 100 time steps.  One time step corresponds to the time that it takes for a passenger to walk the distance between adjacent rows in the airplane.  All passengers walk at the same rate.  Other distributions from which the luggage loading times are selected, such as a Gaussian or exponential distribution, were studied by \citet{steffen} and are not pursued here.  Moreover, the use of different distributions did not significantly alter those results.  The values of these various model parameters are chosen in order to compare effectively the results presented here with those in \citet{steffen}.

As the passengers work their way down the aisle of the airplane they adhere to the following rules: 1) a person will not begin to move unless there are two spaces between them and the person in front of them---a space being equal to a row, 2) if a person is moving, then they will occupy any empty space in front of them prior to stopping (thus, the passengers bunch-up as they come to a halt), 3) passengers require one space either in front of or behind them in order to load their luggage, and 4) passengers only load their luggage into the bins above their assigned row.  I do not account for the time that it takes for a passenger to slide past someone in the aisle or middle seats in order to get to the window seat.  Such a delay would be straightforward to incorporate, but would not serve any significant purpose here.

When a passenger sits down, the energy of their seat is changed to positive infinity (causing the probability that that particular seat will be chosen again to vanish) and the partition function is recalculated for subsequent passengers.  Once a passenger has chosen which seat to occupy (upon entering the aircraft) they do not alter that decision unless their selected seat becomes occupied in the mean time.  In this event, the passenger chooses another seat based upon the partition function that results from using only the seats between their current row to the back of the airplane.  Thus, if a person's seat is taken, they will select a seat somewhere in front of them.

The exception to this rule is if an intended seat becomes occupied and there are no other available seats to the rear of the airplane.  In this case the passenger will reverse his direction and proceed towards the front of the cabin.  Unlike the forward moving case, the backwards moving passenger will sit in the first row that they come to which has an empty seat (again using the partition function for that particular row to select their seat).  These final rules only affect a small fraction of the total passengers; the important idea in this model is that each time there is a decision to be made regarding seat selection, the seat is chosen according to the precepts of statistical mechanics.  While this passenger movement protocols given here may not match exactly what occurs in practice, I assert that the assumptions applied in this study are reasonable.  Moreover, the overall order that the seats become filled is largely independent of these rules; similar seat selection occurs if the passengers are instantly teleported to their seats (though the airplane fills much faster in such a scenario).

\section{Results for full flights\label{results}}

Here I report the results that this model gives for the time required to completely fill the aircraft.  I discuss the effects of changes to the various parameters including the mean time required by passengers to load their luggage.  Then, I discuss the effects of changing the slope of the linear trend in the energies.  Finally, I present some of the effects of changing the different energies for the seats within a given row.  However, I do not give a complete exploration of the effects of changing the energies of the individual seats---instead focusing on a few cases that identify the primary effects of changing these parameters.  Also, I do not explore changes to the airplane geometry, though some discussion of that issue can be found in \citet{steffen}.

\subsection{Differing parameter values\label{parameters}}

\begin{figure}
\begin{center}
\includegraphics[width=\fsize]{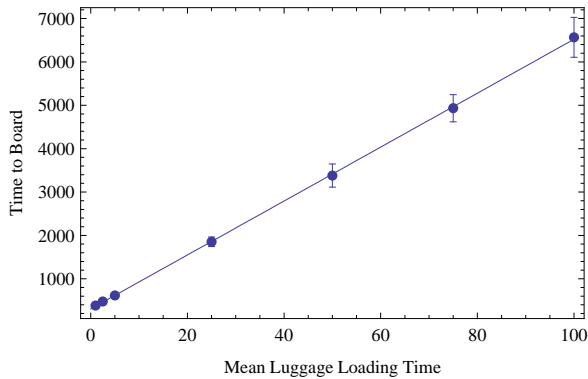}
\caption{A plot of the time required to fill the aircraft for different values of the mean luggage loading times.  These data are generated from 100 realizations of the boarding process, including a reassignment of the luggage loading times for each passenger.  An individual point represents the mean and the error bar is the standard deviation of the resulting distribution of boarding times for the sample.}
\label{timeplot}
\end{center}
\end{figure}

The time required for the all of the passengers to board the aircraft scales linearly with the time that they require to load their luggage.  Figure \ref{timeplot} shows the time required to fill the aircraft for different values of the mean luggage loading times.  Recall that the loading times are assigned to the passengers from a uniform distribution.  These data are generated from 100 realizations of the boarding process (including a reassignment of the luggage loading times) and the error bars represent the standard deviation of the resulting distribution of boarding times.

The best fitting line for these data is
\begin{equation}
\text{Boarding Time} = 62.2\tau + 306
\label{boardtime}
\end{equation}
where $\tau$ is the mean time required to load one's luggage.  The only regime where this relationship deviates from linear is when $\tau$ is much less than the amount of time required by a passenger to walk the length of the airplane.  While this change from the linear behavior is very small, it motivates the use of uniform weights for the data when fitting the line instead of inverse variances.  This is because there is less spread in the distribution of final boarding times for shorter $\tau$ which results in an inappropriate fit for the points with larger $\tau$.  Thus, equation \ref{boardtime} is essentially a least-squares fit as though the points had equal uncertainties.

\begin{figure}
\begin{center}
\includegraphics[width=\fsize]{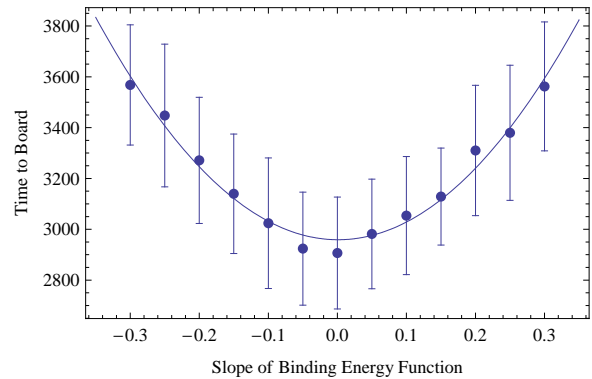}
\caption{The time required to board the airplane as a function of the slope of the linear trend in the energies of the different rows of seats.  The error bars represent the standard deviation of the distributions of the airplane boarding times.  The minimum occurs near a slope of zero.}
\label{slopeplot}
\end{center}
\end{figure}

Another modification that affects the time to board the airplane is to change the slope of the linear trend in the energies of the seats.  The fiducial model has a slope of $+0.25$eV per row, increasing the energies towards the back of the aircraft.  Figure \ref{slopeplot} shows the effect of changing this slope while keeping all other parameters fixed.  The resulting data are best fitted (again using uniform weights) by the equation
\begin{equation}
\text{Boarding Time} = 2960 - 14.7 s + 7100 s^2
\end{equation}
where $s$ is the slope of the linear trend in the energies of the seats.

We see from this figure that the airplane boards faster if there is no trend (or nearly so).  This is because, if there is some preference for one portion of the aircraft over another, then the passengers will tend to pool in that area as they prepare to sit.  This causes an inefficient use of the aisle where the passengers must stand to load their luggage.  When there is no preference for the front or back of the aircraft, then the passengers will more readily spread themselves out along the length of the aisle and, in general, more passengers can load their luggage simultaneously.  This result is similar to what would happen if the temperature of the passengers were increased; the preference for a particular portion of the cabin would become less important and the passengers would spread out more along the aisle.

Altering the energies along a row of seats (the cross section) also affects the boarding times.  For example, if the sign of the energies in a row are changed, such that the middle seats are preferred over the window and aisle seats, then the mean is very near that of the fiducial model, changing from 3380 counts to 3450.  On the other hand, if each seat in a row has the same energy then the airplane boards more slowly---the mean shifting to 3800 counts.  This is due to the fact that, coupled with the linear trend, passengers are more likely to sit in the same portion of the aircraft.  The effect of having different energies for the different seats in a row is that, to some extent, the airplane will fill in phases with the aisle seats largely filling first (for the fiducial model), then the window seats, then the middle seats.  This again is a more efficient use of the aisle of the aircraft since passengers will be stowing their luggage along a larger portion of the length of the aircraft.  This allows the passengers to load in parallel rather than concentrating in a small region of the cabin and loading in serial.

\subsection{Comparison with other boarding methods\label{compare}}

If all of the seats in the entire aircraft have the same energy (or if the model is taken to the high temperature limit), then the boarding is nearly equivalent to random boarding with assigned seats.  Indeed, the mean of an ensemble of realizations of this case is 2900 ($\pm 220$) which is very close to the mean of 2846 reported in \citet{steffen} for this case.  Most of the difference can be accounted for by passengers who must reverse their direction and return to the front of the aircraft in order to find an empty seat---something that would not occur with the assigned seating model.

The overall boarding times that are obtained with the FFA model are roughly the same as the best ``practical'' boarding strategies\cite{steffen} with assigned seats such as boarding window seats first, then middle seats, then aisle seats.  While a direct quantitative comparison between these models is not warranted without proper calibration, the assumptions that are employed here give results that are in good qualitative agreement with those in \citet{steffen} and in industry practice.  Thus, the model described here can be used to qualitatively compare the benefits of different boarding procedures including FFA boarding and methods which employ assigned seats.

\section{Partially filled aircraft\label{fermigas}}

To this point we have only looked at the case of a fully loaded airplane where the final ``state'' of the entire airplane is predetermined.  However, in the event that the flight is only partially filled, statistical mechanics can also predict the likely seating arrangements of the passengers after the boarding process has completed.  Or, it can be used to predict the distribution of seated passengers at various times during the boarding of a full flight.

In this case, it is more straightforward to consider the passengers as a fermi gas (which is true for all passengers except adults carrying infants in their lap) and use the Fermi-Dirac distribution to determine the probability that a particular seat is occupied at the time of interest.  The occupancy of a seat $n$ with energy $\epsilon$ is given by
\begin{equation}
n = \frac{1}{e^{(\epsilon - \mu)/T} + 1}
\label{fermidirac}
\end{equation}
where $\mu$ is the chemical potential of the system in question.  Here, the chemical potential characterizes the energy of the seat that is likely to be occupied by the next passenger to board the airplane.

\begin{figure}
\begin{center}
\includegraphics[width=0.4\textwidth]{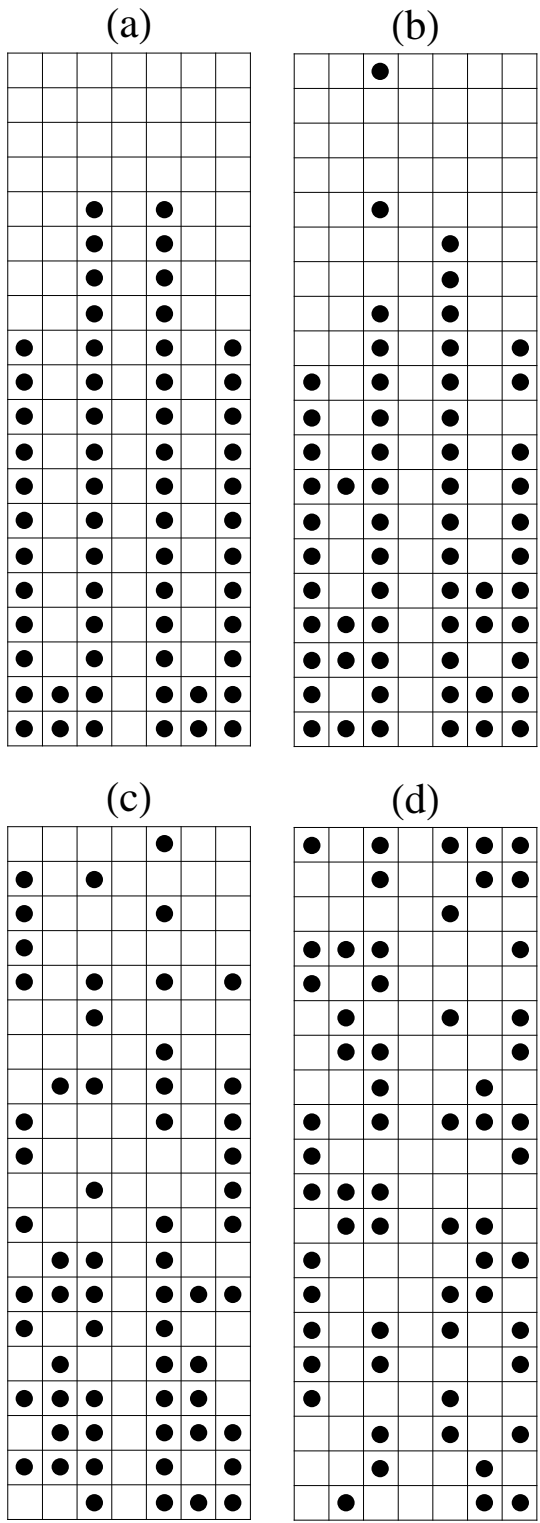}
\caption{Final seating arrangement for a half-full airplane for different values of the temperature: (a) is for a temperature of $|\bar{\epsilon}|/1000 \simeq 0$, (b) is for a temperature of $| \bar{\epsilon} |/10$, (c) is for a temperature of $|\bar{\epsilon}|$, and (d) is for a temperature of $10|\bar{\epsilon}|$ where $\bar{\epsilon}$ is the mean energy of the seats in the nominal airplane model.  The dots represent seated passengers.}
\label{partfilled}
\end{center}
\end{figure}

The chemical potential depends upon both the number of passengers in the aircraft and the temperature.  At zero temperature the first unoccupied seat gives the chemical potential and is equal to the fermi energy.  As the temperature increases, the chemical potential may be found by recognizing that the number of passengers is fixed and solving
\begin{equation}
N = \sum_{i} \frac{1}{e^{(\epsilon_i-\mu)/T}+1}
\end{equation}
for $\mu$ at the desired temperature.  The result of this calculation is not included here but the calculation itself is suitable as an undergraduate level computational problem.  Figure \ref{partfilled} shows the final seating arrangements for the fiducial airplane model where a different temperature is used for each of the realizations.  This figure shows how the seating distribution changes from a degenerate arrangement to a random arrangement as the temperature increases.

\begin{figure}
\begin{center}
\includegraphics[width=\fsize]{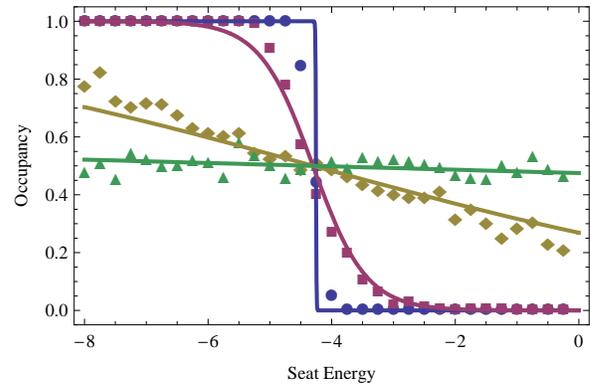}
\caption{Seat occupancies (the fraction of the time that a seat with a given energy is occupied) as a function of energy for 100 realizations of the boarding process with various values for the passenger temperature.  The temperature values match those of figure \ref{partfilled}: $|\bar{\epsilon}|/1000 \simeq 0$ (circles), $| \bar{\epsilon} |/10$ (squares), $|\bar{\epsilon}|$ (diamonds), and $10|\bar{\epsilon}|$ (triangles) where $\bar{\epsilon}$ is the mean energy of the seats.  The curves are theoretical predictions using the Fermi-Dirac distribution.}
\label{datatheory}
\end{center}
\end{figure}

Figure \ref{datatheory} shows the results of 100 realizations of the passenger boarding algorithm at various temperatures and compares those results to the corresponding prediction of the Fermi-Dirac occupancy distribution.  Here there are are only 60 passengers, the mean boarding time is 15 time steps (to lessen the computational expense), and the temperature is set to the stated values, otherwise the input parameters match those of the fiducial model.  Of particular interest is that the Boltzmann-type decisions of the individual passengers coupled with the the fact that passengers don't move once seated recovers the Fermi-Dirac distribution; the largest discrepancy being for very low temperatures where both distributions exhibit mathematical pathologies and where the discrete nature of the allowed seat energies becomes more important (the chemical potential is generally not equal to one of the allowed energy values so the distribution crosses a value of 1/2 between the energy levels).

Note that while the chemical potential is often introduced in statistical mechanics classes by allowing the environment to exchange particles with the system (the grand canonical ensemble), this is not a necessary condition for its use.  A discussion on the relationship between the chemical potential, temperature, energy, and number of particles in a system can be found in \citet{kubo}.  In addition, when particles can be exchanged with the environment, is not necessary that the system and its environment be physically distinct (see, for example \citet{schroeder}).  For the system here, with a fixed number of particles in thermal equilibrium with a heat reservoir, the chemical potential is implicitly defined by the temperature and the number of particles.  It characterizes the expected energy that the system would gain (or lose) if an additional passenger were injected (or removed) regardless of whether or not it is physically possible to do so.

\section{Conclusions\label{discussion}}

In this article I have shown that the decisions made by the passengers in the FFA airplane boarding process may be effectively modeled using the principles of statistical mechanics, namely that the desirability of a particular seat can be expressed as an energy and the probability of someone choosing a particular seat can be estimated with Boltzmann statistics.  One can also use Fermi-Dirac statistics to identify the probability that a particular seat will be occupied at any time during the boarding process.  Airplane boarding is a somewhat unusual application of statistical mechanics, yet it may be a valuable pedagogical example for an undergraduate course since it is both novel and something that many students and faculty have experienced first hand.  Moreover, showing such non-standard applications of physical theory will likely prove valuable to students since most will ultimately find employment in industrial settings where ``outside-the-box'' thinking will be beneficial.

Some additional modifications to this model which would affect the boarding process, such as how the energies of a particular portion of the airplane change as a passenger sits in a given seat, how crowding in the aisle affects the local energies (or the temperature of the waiting passengers), or the increased time required by someone sitting in a window seat compared to an aisle seat might improve the model's accuracy.  However, here those changes would only serve to complicate the facts that statistical physics is an appropriate framework to model FFA passenger boarding and that airplane boarding is an interesting application of undergraduate level physics.

Another modification would be to allow different passengers to enter the aircraft at different temperatures (indicating different arrival histories at the gate).  The effect here of each passenger being at a different temperature upon entering the airplane would be to change their individual decisions regarding their seat selection.  Since this model does not have passengers changing seats, the seat selection process for most passengers is fixed by the available seats at the time that they enter the airplane (the exception is when two passengers inadvertently select the same seat).  Thus, while this scenario may be challenging to set up in a traditional thermodynamic system, it may yet be appropriate here where the temperature reflects the temperament of the passenger in question.  Thus, while some passengers may not care where they sit (high temperature), others at a lower temperature would.  Identifying where the uniform temperature approximation might fail would be an interesting result.  Also, students could use this model to calculate, for example, the heat capacity of a partially filled airplane or perhaps identify the temperature of the passengers based upon the occupied seats.

In addition to passenger boarding, statistical mechanics might be the proper framework to model passengers leaving the airplane.  Here the passengers, who are essentially frozen in a crystalline structure, would sublimate and escape into the environment, the terminal.  There is a ``latent heat'' associated with the transition which is characterized by the time that is required for a passenger to retrieve his or her luggage.

Beyond the classroom, an appropriate model for FFA boarding is valuable to people in the airline industry since the insights gained by a proper model can be used to make decisions regarding a company's boarding policies and practices or even where they may store the blankets and pillows (presumably above seats that are not likely to be occupied).  While correct calibration and sufficient detail are important, even the basic principles that can be extracted from the model discussed here, such as the effect of the slope in the linear trend or the effect of more uniform energies in a particular row can effectively inform the interested party.  Moreover, as seen in figure \ref{slopeplot}, the fastest boarding times occur when the boarding process is nearly random.  This indicates that raising the ``temperature'' (or temperament) of the passengers would tend to increase the speed of the boarding process.  While changing this passenger temperature has components which lie outside the control of an airline company, some influence may still exist whereby passengers generic predisposition to certain seats could be reduced, effectively making all the seats, or at least all of the rows, equivalent---something that could further improve upon this already successful boarding strategy.

\section*{Acknowledgements}

J. Steffen acknowledges the generous support of the Brinson Foundation.  Fermilab is supported by the U.S. Department of Energy under contract No. DE-AC02-07CH11359.


\begin{thebibliography}{5}

\bibitem[Metropolis et al.(1953)]{metro} N. Metropolis, A.~W. Rosenbluth, M.~N. Rosenbluth, A.~H. Teller, and E. Teller., ``Equations of State Calculations by Fast Computing Machines,'' Journal of Chemical Physics, 21(6):1087-1092, (1953).

\bibitem[Press et al.(2002)]{press} W.~H. Press, S.~A. Teukolsky, W.~T. Vetterling, and B.~P. Flannery, \textit{Numerical Recipes in C}, Cambridge University Press, USA, (2002).

\bibitem[Bachmat et al.(2006)]{bachmat} E. Bachmat, D. Berend, L. Sapir, S. Skiena, and N. Stolyarov, ``Analysis of airplane boarding via space-time geometry and random matrix theory,'' Journal of Physics A, 39:L453-459, (2006).

\bibitem[Steffen(2008)]{steffen} J.~H. Steffen, ``Optimal boarding method for airline passengers,'' Journal of Air Transport Management, Accepted, (2008), arXiv:0802.0733.

\bibitem[Van Landeghem \& Beuselinck(2002)]{belgians} H. Van Landeghem and A. Beuselinck, ``Reducing passenger boarding time in airplanes: A simulation based approach,'' European Journal of Operational Research, 142:294-308, (2002).

\bibitem[van den Briel et al.(2005)]{vandenbriel} M.~H.~L. van den Briel, J.~R. Villalobos, G.~L. Hogg, T. Lindemann, and A. Mul\'e, ``America West Develops Efficient Boarding Strategies,'' Interfaces, 35(3):191-201, (2005).

\bibitem[Demerjian(2006)]{wired}  D. Demerjian, ``Airlines Try Smarter Boarding,'' Wired Magazine Online, May 9, 2006.

\bibitem[Kubo(1964)]{kubo} R. Kubo, in cooperation with H. Ichimura, T. Usui, N. Hashitsume, ``Statistical Mechanics: an Advanced Course with Problems and Solutions,'' North-Holland, Amsterdam, (1965, 7th ed. 1988).

\bibitem[Schroeder(1999)]{schroeder}  D. V. Schroeder, ``An Introduction to Thermal Physics,'' Addison-Wesley Publishing Company, San Francisco, CA, (1999).

\end{thebibliography}
\end{document}